\documentclass[10pt,conference]{IEEEtran}
\IEEEoverridecommandlockouts
\usepackage{cite}
\usepackage{amsmath,amssymb,amsfonts}
\usepackage{algorithmic}
\usepackage[table,xcdraw]{xcolor}
\usepackage{graphicx}
\usepackage{textcomp}
\usepackage{xcolor}
\usepackage{booktabs}
\usepackage{soul}
\usepackage{hyperref}
\usepackage{multirow}
\usepackage{tcolorbox}
\usepackage{enumitem}
\usepackage{amsthm}
\usepackage{balance}
\theoremstyle{definition}
\newtheorem{definition}{Definition}%[section]

\def\BibTeX{{\rm B\kern-.05em{\sc i\kern-.025em b}\kern-.08em
    T\kern-.1667em\lower.7ex\hbox{E}\kern-.125emX}}

\newboolean{showcomments}
\setboolean{showcomments}{true}         
\ifthenelse{\boolean{showcomments}}
  {\newcommand{\nb}[2]{
  \fbox{\bfseries\sffamily\scriptsize#1}
     {\sf\small$\blacktriangleright$\textit{\textcolor{red}{#2}}$\blacktriangleleft$}
   }
  }
  {\newcommand{\nb}[2]{}
   
  }

\newcommand{\COMMENT}[1]{}

\makeatletter
\newcommand{\thickhline}{%
    \noalign {\ifnum 0=`}\fi \hrule height 1pt
    \futurelet \reserved@a \@xhline
}
\newcolumntype{"}{@{\hskip\tabcolsep\vrule width 1pt\hskip\tabcolsep}}
\makeatother

\begin{document}

\title{When and Why Test Generators for Deep Learning Produce Invalid Inputs: an Empirical Study\\
\thanks{This work was partially supported by the H2020 project PRECRIME, funded under the ERC Advanced Grant 2017 Program (ERC Grant Agreement n. 787703).}
}

\author{\IEEEauthorblockN{Vincenzo Riccio}
\IEEEauthorblockA{\textit{Università della Svizzera italiana} \\
Lugano, Switzerland \\
https://orcid.org/0000-0002-6229-8231}
%vincenzo.riccio@usi.ch}
\and
\IEEEauthorblockN{Paolo Tonella}
\IEEEauthorblockA{\textit{Università della Svizzera italiana} \\
Lugano, Switzerland \\
https://orcid.org/0000-0003-3088-0339}
%paolo.tonella@usi.ch}
}

\maketitle

%!TEX root = 0_main_TIG_validity.tex
\begin{abstract}
% Testing DL. 
Testing Deep Learning (DL) based systems inherently requires large and representative test sets to evaluate whether DL systems generalise beyond their training datasets.
%
% TIGs
Diverse Test Input Generators (TIGs) have been proposed to produce artificial inputs that expose issues of the DL systems by triggering misbehaviours.
%
% Input Validity. 
Unfortunately, such generated inputs may be invalid, i.e., not recognisable as part of the input domain, thus providing an unreliable quality assessment.
%
% Validators
Automated validators can ease the burden of manually checking the validity of inputs for human testers, although input validity is a concept difficult to formalise and, thus, automate.

In this paper, we investigate to what extent TIGs can generate valid inputs, according to both automated and human validators.
We conduct a large empirical study, involving 2 different automated validators, 220 human assessors, 5 different TIGs and 3 classification tasks.
Our results show that 84\% artificially generated inputs are valid, according to automated validators, but their expected label is not always preserved. Automated validators reach a good consensus with humans (78\% accuracy), but still have limitations when dealing with feature-rich datasets.

\end{abstract}

%!TEX root = 0_main_TIG_validity.tex
\section{Introduction} \label{introduction}

Deep Learning (DL) systems have become paramount to Software Engineering (SE), due to their ability to learn how to solve complex tasks from training data~\cite{Manning-IIR-2008}. 
Such ability stands them apart from other software systems, but is a double-edged sword since their behaviour cannot be foreseen by source code analysis, as happens for traditional software~\cite{RiccioEMSE20}. 

A best practice to assess the quality of DL systems is to partition the original labelled dataset into training, validation, and test set~\cite{lecun2015deep}; the system is evaluated on the test set through performance metrics, such as the classification accuracy, i.e., the percentage of correct predictions. 
However, obtaining large and representative test sets along with the corresponding correct labels is often unfeasible or extremely onerous. 
%Therefore, DL systems should be carefully tested with appropriate techniques that generate data beyond the systems' datasets~\cite{zhangTSE22, RiccioEMSE20}. 

In the SE literature, several Test Input Generators (TIGs) have been proposed to produce artificial inputs, which are used to evaluate whether a DL system generalises beyond the initially considered dataset of inputs, since often such dataset is not fully representative of the scenarios that will be experienced in the field~\cite{zhangTSE22, RiccioEMSE20}. 
Most TIGs target image classifiers and generate inputs by adopting different strategies, i.e., (1) pixel-level perturbation of existing images~\cite{PeiCYJ17, guo2018dlfuzz}; (2) manipulation of the input representation provided by generative DL models~\cite{kang2020sinvad, DunnISSTA21}; (3) modification of domain-specific model representations~\cite{RiccioFSE20, ZohdinasabISSTA2021}. 
%
%TIGs' goal is to generate misbehaviour-inducing inputs, i.e., inputs for which the label predicted by the DL system deviates from the expected one. 
The goal of TIGs is triggering misbehaviours, i.e., deviations of the DL system prediction from the expected label.
However, the generated test inputs may be invalid and provide an unreliable quality assessment~\cite{DolaICSE21}.

The concept of \textit{input validity} is crucial for software testing, since it indicates that the considered inputs are at once \textit{relevant} and \textit{meaningful} (i.e., they belong to the input domain, hence they are interpretable and can be handled by the software system under test). 
Traditional software is typically equipped with preconditions to check the  validity of the inputs. For instance, a function to compute the square root of a positive number is supposed to raise an exception if invoked with a negative number. 
Hence, a valid input for traditional programs is one that satisfies their preconditions. 
Instead, DL systems accept any input conformant with the input tensor shape, regardless of its validity. 
For instance, a DL system that classifies pictures of animals, when exercised with the image of a car, will still return the label of an animal, without raising exceptions. 
Therefore, defining input validity for a DL system is way more difficult than for traditional software. In this paper we adopt the following definition:
\begin{definition}[\textbf{Input Validity for DL}]\label{def:validity}
\textit{A valid input for a DL system is one that is recognisable by human domain experts in the input domain, i.e., an input to which a human can confidently assign a %unique 
label taken from the input domain.}
\end{definition}
Therefore, a picture of a car is not a valid input for DL classifiers of animal images.\footnote{
It should be noticed that \textit{realism} and \textit{validity} are different notions: a digitally modified image of a dog might remain recognisable as an animal for humans, but it might look extremely unrealistic.}
%
%Such an image would be valid but not realistic as no real-world animal would naturally look as in such image.
%
Hence, TIG outputs should be validated by checking that the generated images are valid, i.e. they belong to the specific domain of the classification/regression task. 
Since manually checking the generated test inputs can easily become a burden for human testers, recent works~\cite{DolaICSE21, StoccoICSE19} proposed automated validity assessment techniques. %based on the usage of Variational Auto-Encoders (VAEs) as out-of-distribution detectors~\cite{yang2021generalized}.
However, these works did not investigate how much their validity assessment agrees with human judgement. 

Another problem that affects TIGs is the oracle problem, i.e., how to define the expected label of each generated input. 
TIGs address this problem by starting from seeds with known ground-truth labels and by applying small perturbations that are supposed to keep the original labels unchanged. 
%
%However, these mechanisms do not necessarily preserve labels and should also be validated. 

Existing TIGs do not consider the validity of the generated test inputs and assume  the original ground-truth labels are preserved. 
Dola et al~\cite{DolaICSE21} performed a study which shows that TIGs may generate invalid inputs, which influence test adequacy metrics, e.g., neuron coverage~\cite{PeiCYJ17}. 
However, their study was limited to pixel-perturbing TIGs and automated validators. They also did not consider the problem of ground-truth label preservation.
%\vincenzo{they considered only simple classification tasks with small images and few (i.e., 10) classes.}

In this paper, we present a large empirical study evaluating $5$ different TIGs, with respect to their ability to generate valid, misbehaviour-inducing inputs and preserve their ground-truth labels.
We considered both the validity assessment performed by automated validators and humans. Specifically, we applied two existing automated validators from the literature and we involved $220$ human assessors from a crowdsourcing platform. 
Our empirical study is the first to compare multiple TIGs using both automated and human validators, and to assess the assumption of label preservation made by such TIGs.

Our results show that all TIGs can produce, to different extents, valid inputs according to both automated and human validators, but they do not always preserve the ground-truth label of the inputs. 
Moreover, we found a good match between automated validation and human judgement. 
Nevertheless, our study revealed limitations of automated validators when dealing with complex, feature-rich datasets and with simple images that show characteristics different from their original dataset but are still valid for humans.

%To define the value of the expected label for a newly generated, artificial input, i.e., to address the oracle problem, these techniques apply small perturbations to inputs for which the expected label is known, under the assumption that such perturbations will not alter the label.

%TIGs aim at assessing the quality of DL models by producing misbehaviour-inducing inputs,  i.e., inputs for which the predicted label differs from the ground-truth. Let us consider the case of DL image classifiers. The automated generation of valid images to test a DL image classifier is a challenging task, because of the semantic manifold problem~\cite{Yoo2019}: the input space, consisting of all possible pixel value combinations, is huge, while the manifold contained in the input space that is semantically relevant for the given task (e.g., images of animals) is a tiny portion of it. 

%During input generation, TIGs navigate such a complex input space. This implies that artificial inputs produced by TIGs may become invalid, because they do not lie in the semantic manifold anymore. 

%!TEX root = 0_main_TIG_validity.tex
\section{Validity Assessment of Test Inputs for DL} \label{validity}

Let us consider DL image classifiers. 
The automated generation of valid test images is a challenging task because of the semantic manifold problem~\cite{Yoo2019}: the input space, consisting of all possible pixel value combinations, is huge, while the manifold contained in the input space that is semantically relevant for the given task (e.g., images of animals) is a tiny portion of it. 
%%However, classifiers predict and report a label for any of pixel value combination, even meaningless ones (e.g., white noise images).
During input generation, TIGs navigate such a complex input space. This implies that artificial inputs produced by TIGs may become invalid, because they no longer lie in the semantic manifold. 
Hence, TIG outputs should be validated by checking that the generated, artificial images are valid, i.e. they belong to the specific domain of the classification task. 
%Moreover, most TIGs solve the oracle problem by starting from a seed for which the ground truth label is known and by applying small perturbations that are supposed to keep the original ground truth label unchanged. Again, this is not necessarily true and should be also validated. 

Since manually checking automatically generated test inputs can easily become a burden for human testers~\cite{attaoui2022black}, recent  works~\cite{DolaICSE21, StoccoICSE19, StoccoGAUSS20, StoccoJSEP21} proposed automated validity assessment techniques based on the usage of Variational Auto-Encoders (VAEs) as out-of-distribution detectors~\cite{yang2021generalized}.

%VAEs
VAEs are DL models consisting of two sequentially connected  components: the encoder and the decoder~\cite{kingma2013auto}. The \textit{encoder} maps the original input to a so-called \textit{latent space}, consisting of a normal multivariate probability distribution of parameters, represented as the means and variances  of a $z$-dimensional normal distribution. The dimensionality $z$ of the latent space is usually much smaller than the original input space one. Thus, VAEs can be effectively used also for dimensionality reduction. 
The \textit{decoder} maps a $z$-dimensional vector, which represents a sample from the normal multivariate distribution, to an input space vector: it performs the reverse transformation, by reconstructing the input in the original space starting from the encoded vector.

Both the encoder and the decoder are two neural networks that are jointly trained to minimise both the reconstruction error (i.e., the difference between the original inputs from the training set and the VAE's reconstructions), and the Kullback-Leibler divergence between the learned posterior and the true posterior (usually modeled as a unit Gaussian distribution, for each dimension of the latent space)~\cite{an2015variational}. 

%After training, 
When a trained VAE processes an image not represented in the training set (i.e., an out-of-distribution image), it produces a less precise reconstruction than the one obtained for inputs similar to the training data. Such a property of VAEs is leveraged by the two main automated validity assessment techniques proposed in the  literature: Distribution-Aware Input Validation (DAIV)~\cite{DolaICSE21} and SelfOracle~\cite{StoccoICSE19}.

\subsection{Distribution-Aware Input Validation (DAIV)} \label{dola}

%Dola et al.~\cite{DolaICSE21} proposed DAIV, a VAE-based input validator for DL-based image classifiers. In particular, the authors apply DAIV to inputs generated for classification problems (i.e., handwritten digits and house numbers). 
DAIV is a VAE-based input validator for DL image classifiers~\cite{DolaICSE21}.
DAIV is distribution-aware since its VAE is trained on the same data used to train the DL system under test, i.e., both DL models learn the same distribution. Additionally, it requires the availability of an anomaly set, containing examples of out-of-distribution inputs.

DAIV leverages the VAE architecture by An and Cho~\cite{an2015variational}, whose decoder is probabilistic: it returns the reconstructed image in the form of a probability distribution %of values 
(i.e., pixel values are replaced by $\hat{\mu}$ and $\hat{\sigma}$ values). This VAE is trained to maximise the reconstruction fidelity $R$, quantified as the negation of two loss metrics: $loss_m$, which depends on the reconstruction error, and $loss_a$, which measures the variance of the reconstructed distribution. 
\COMMENT{
For a single probabilistically reconstructed pixel value, the loss metrics are defined as follows:

%formula
\begin{eqnarray}
\scalebox{.9}{$loss_m = \sqrt {\frac{\hat{\mu} - x } {\hat{\sigma}}} $}
& & 
\scalebox{.9}{$loss_a = \log 2 \pi {\hat{\sigma}} $}
\end{eqnarray}

\noindent
where $x$ is the original pixel value.
Assuming the input image has width $W$, height $H$ and $C$ channels (so, in total $W \times H \times C$ pixel values), the reconstruction fidelity $R$ across all pixels is defined as follows:

\begin{equation}
\scalebox{.9}{$\displaystyle R = -1/2  \sum_ {i = 1, j = 1, c = 1}^{W, H, C}{loss_a[c][i,j] + loss_m[c][i,j]}$}
\end{equation}
}

After training, the VAE will exhibit higher $R$ values for inputs belonging to the training data distribution (i.e., nominal data), with respect to out-of-distribution inputs (i.e., anomalies). For this reason, the authors compute fidelity estimations both on nominal and on anomalous data, in order to identify a threshold  above which VAE inputs can be considered valid. To find the optimal threshold, the identified nominal and anomalous datasets are tested against multiple threshold values. The candidate threshold that produces the highest F-measure is chosen as final threshold for the input validator, since it best separates the two datasets (as shown in ~\autoref{fig:tshds}~(a)). As a result, the artificial inputs generated by a TIG with $R$ lower than the optimal threshold are classified as invalid by the distribution-aware input validator. 
%\vincenzo{here i can add an image with the histograms for the two different datasets and the threshold and/or a plot of the F-measures per threshold as shown in out meeting} \paolo{I like the first plot, but only if space is left, so let's wait and see if any space is left in the paper}
%
The choice of the anomalous dataset is critical for this technique since it affects the threshold value and, thus, the validity assessment, e.g., nominal and anomalous datasets with overlapping distributions cause a high rate of false alarms~\cite{weiss2022forgotten}. The authors suggest choosing a dataset that (1) has the same input size as the nominal dataset, and (2) encodes different categories from the ones of the nominal dataset (e.g., for a classifier of animal images, images of cars). In this way,  nominal and anomalous datasets are  likely to have completely different distributions.

\subsection{SelfOracle}

\begin{figure}[t!]
	\center \includegraphics[width=180pt]{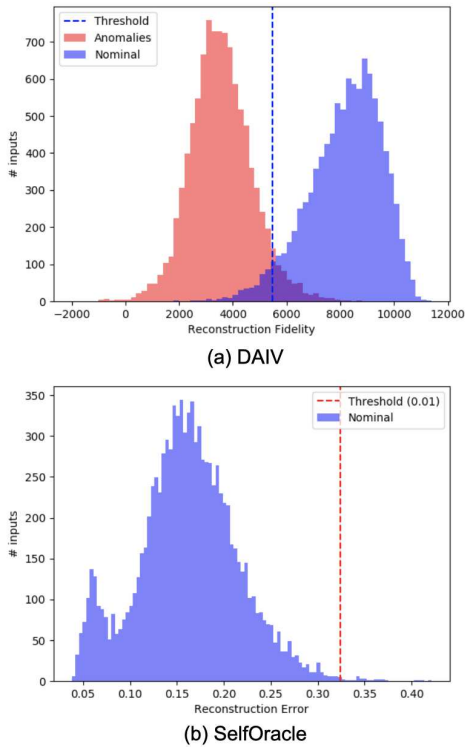}
        \caption{Validation thresholds defined by automated validators}
        	\label{fig:tshds}
\end{figure}

SelfOracle is a VAE-based anomaly detector originally proposed for autonomous driving systems~\cite{StoccoICSE19}. This technique also exploits a VAE trained on the same data used to train the DL system under test, but it does not require the availability of any anomaly set. 
%
%SelfOracle's VAE reconstructs the images captured by the on-board car's camera to detect unexpected driving scenarios and prevent potentially fatal misbehaviours such as collisions or out-of-bound episodes. 
%
The authors adopted VAEs trained to minimise the reconstruction error, where the reconstructed image is obtained from a deterministic decoder. 
After training, the SelfOracle's VAE will have a higher reconstruction error on invalid data (out-of-distribution), whereas it will return better reconstructions for valid data.

SelfOracle leverages probability distribution fitting to find the threshold that discriminates nominal (valid) inputs from anomalous (invalid) ones. In particular, it fits a Gamma distribution to the training data through the maximum likelihood estimation method. Probability distribution fitting provides a statistical model for the reconstruction errors returned by the VAE for nominal data. This statistical model can be used by the tester to configure the threshold for the VAE's reconstruction error that brings the desired rate of false alarms (i.e., inputs that belong to the nominal dataset but are classified as invalid). 
For instance, a 1\% false alarm rate means choosing the threshold value that splits the Gamma distribution into two parts, the rightmost one (associated with higher reconstruction errors) having an integral equal to 0.01 (see ~\autoref{fig:tshds} (b)).
Inputs with reconstruction error above such threshold are considered anomalous (invalid), while inputs below it are considered nominal (valid). 
\section{Test Input Generation for DL} \label{tig}

\begin{figure}[t!]
	\center \includegraphics[width=230pt]{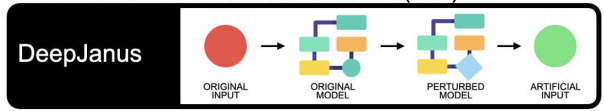}
        \caption{Main approaches to input generation}
        	\label{fig:tigs}
\end{figure}

Several TIGs have been proposed to generate artificial, misbehaviour-inducing inputs~\cite{zhangTSE22, RiccioEMSE20}, i.e., inputs for which the label predicted by the DL system deviates from the expected one. To define the value of the expected label for a newly generated input, i.e., to address the oracle problem, these techniques apply small perturbations to inputs for which the expected label is known, under the assumption that such perturbations will not alter the label.

% other objectives
TIGs may have additional objectives to exposing misbehaviours, such as optimising DL-specific coverage metrics (e.g., neuron~\cite{PeiCYJ17} or surprise coverage~\cite{KimFY19}). The idea behind these coverage objectives is that we can gain higher confidence on the DL system behaviour when all input partitions induced by the coverage criterion are exercised without exposing misbehaviours or, when misbehaviours are indeed exposed, that the newly generated inputs can help developers improve the DL system. To avoid generating too similar inputs, some TIGs consider also input diversity as a test objective~\cite{RiccioFSE20}.
%They may aim to trigger differential behaviours, i.e. generate a single input that triggers different behaviours on multiple models or a pair of similar inputs that trigger different behaviours of the same model. 
%targeted misprediction (FP)

% algorithms and access: black box, white box
TIGs may have white box access to the DL system under test, which allows them to acquire information such as the neurons' activation values. Otherwise, TIGs leverage information about the inputs and from the output (e.g., \textit{softmax}) layer to guide input generation (black box access). 

% Input representation and manipulation
The way inputs are represented and perturbed is crucial for their validity. In the following, 
%we focus on images as inputs and 
we present the three main approaches to handle input images, depicted in \autoref{fig:tigs}:  (1) Raw Input Manipulation, (2) Generative DL models, and (3) Model-based Input Representation. For each of these approaches we list representative tools from the literature, in particular those that we considered in our empirical study. The respective objectives, which are optimised in the test generation process, are shown in Table~\ref{tab:tig-features}.

\begin{table}[]
\caption{Characteristics of the considered test input generators} \label{tab:tig-features}
\begin{tabular}{llll}
%\toprule
\thickhline    
\textbf{Tool}                  & \textbf{Approach} & \textbf{Objectives}                                                                  & \textbf{Access}    \\
\midrule
DeepXplore            & RIM   & \begin{tabular}[c]{@{}l@{}}Misbehaviours, \\ Neuron Coverage\end{tabular}   & White-box \\
\midrule
DLFuzz                & RIM   & \begin{tabular}[c]{@{}l@{}}Misbehaviours, \\ Neuron Coverage\end{tabular}   & White-box \\
\midrule
Sinvad                & GDLM  & \begin{tabular}[c]{@{}l@{}}Misbehaviours, \\ Surprise Coverage\end{tabular} & \begin{tabular}[c]{@{}l@{}}White-box, \\ Black-box\end{tabular} \\
\midrule
Feature Perturbations & GDLM  & Misbehaviours                                                               & Black-box \\
\midrule
DeepJanus             & MIM   & \begin{tabular}[c]{@{}l@{}}Misbehaviours, \\ Frontier, Diversity\end{tabular}         & Black-box\\
%\midrule
\thickhline    
\end{tabular}
\end{table}

\subsection{Raw Input Manipulation (RIM)}

RIM approaches modify an image in the original pixel space to generate a new input, by perturbing pixel values. These approaches generally start from a seed input for which the label is known and the system performs correctly. Then, they manipulate the input by adding an input mask of perturbations, which are as imperceptible as possible, while at the same time expose a system misbehaviour.

DeepXplore~\cite{PeiCYJ17} belongs to RIM approaches. It adds effects to the seed image to trigger misbehaviours, including: (1) \textit{occlusion}, consisting of a small patch of random pixels in a predefined position; (2) \textit{light}, uniformly increasing/decreasing the pixel values to simulate light exposure; (3) \textit{blackout}, increasing the pixel values of random squares. These perturbations are not totally random, as they are crafted to optimise neuron coverage of the DL system under test.

Another RIM technique is DLFuzz~\cite{guo2018dlfuzz}, which adds noise to the seed image to increase the probability of misbehaviour, estimated from the DL system's output layer. Similarly to DeepXplore, DLF also maximises neuron coverage.

RIM techniques mainly generate adversarial inputs, i.e., they simulate perturbations performed by a malicious attacker to induce a misbehaviour through imperceptible changes of the input. Such images are not necessarily representative of those experienced in the field, since it is difficult to observe similar corruptions in the real world, if not due to the malfunctioning of the camera. So, these techniques are more relevant for security testing than for functional testing.

%RIM approaches are fast and effective since they easily produce misbehaviours. 
%However, they apply small perturbations to inputs for an important reason: the input space is huge and if they modify too much an input, it can loose meaning

Since RIM techniques can modify only existing inputs, their performance depends on the quality of the initial seeds. Another drawback of these techniques is that they cannot explore the original input space at large. %In fact, the input space is huge, while only small portions of it are meaningful. 
By applying only small perturbations to existing valid inputs, these techniques may remain confined within regions reachable from the seeds, leaving many other valid input regions untested.

\subsection{Generative DL models (GDLM)}

Generative DL models can reconstruct the underlying distribution of the input data and exploit this knowledge to generate new inputs. Widely used generative models are Variational AutoEncoders (VAEs)~\cite{kingma2013auto} and Generative Adversarial Networks (GANs)~\cite{goodfellow2014generative}. VAEs are trained with the objective of minimizing the reconstruction error measured when a training image is encoded into a latent space vector and is then decoded into an image. GANs add a discriminator to the reconstruction process, introducing a form of competitive learning in which the discriminator learns to distinguish real from artificially generated images, while the encoder-decoder network (i.e., the generator) learns also how to fool the discriminator.

GDLM techniques operate on a ``latent'' input space that is much smaller than the original input domain, often mapped to a $z$-dimensional standard normal distribution. VAE decoders or GAN generators then transform latent vectors into images with the dimensionality of the original input space.

Sinvad~\cite{kang2020sinvad} is a GDLM technique that directly perturbs the latent vector. In particular, it adds a random value sampled from the standard normal distribution to a single element of the latent vector.
The latent space exploration performed by Sinvad is guided either by the probability of misbehaviours, estimated from the output softmax layer, or by surprise coverage~\cite{KimFY19}, to be  maximised, under the assumption that surprising inputs are also likely to expose misbehaviours. In order to quantify the degree of surprise of an input, Sinvad relies on the distances between activation vectors or on kernel density estimation. 
%In the former case, an input has high surprise if the associated activation vector is far from that of the closest input of the same class, which in turn is close to that of another input from a different class. In the latter case, an input has high surprise if its probability of occurrence is low according to kernel density estimation. Regardless of the adopted surprise metric, the range of surprise values is partitioned into bins and coverage is defined w.r.t. such bins. \vincenzo{the Sinvad version we used is guided by softmax, not surprise. We can either remove this explanation, or say that Sinvad \textbf{can} be guided by surprise.}

The Feature Perturbations technique~\cite{DunnISSTA21, dunn2020evaluating} injects perturbations into the output of the generative model's first layers, which represent  high level features of images. Perturbations are 
%of the feature layer outputs is 
guided by a metric quantifying the distance from a misclassification, still computed from the softmax outputs.

Differently from RIM techniques, generative DL models operate on a significantly reduced input space and produce realistic inputs that can be quite distant from the inputs in the original dataset. The quality of the inputs generated by GDLM techniques strictly depends on the quality of the training set and of the generative model.

\subsection{Model-based Input Manipulation (MIM)}

MIM approaches generate test inputs by relying on a model representation of their domain. These approaches are reminiscent of classic model driven engineering. The original input is transformed into an  input model instance which abstracts its main features. The manipulation is performed on the model, which is finally transformed back to the original format~\cite{Larman1997}.

DeepJanus~\cite{RiccioFSE20} is a MIM approach that was applied to classification and regression problems, including handwritten digit image recognition (classification), self-driving (regression) and eye-gaze prediction (regression).
For image classification, DeepJanus abstracts a model instance from the bitmap, obtaining a Scalable Vector Graphics (SVG) representation of handwritten digits. In this way, DeepJanus can directly modify the model instance (i.e., the control parameters of the SVG representation) and, then, transform it back to the original input format by means of a rasterisation operation.

MIM approaches operate on a reduced input space (e.g., the control parameters of the SVG representation) and, with proper model constraints, ensure realism of the generated inputs. However, the main limitation of these approaches is that they require the existence of a good-quality model representation for the given input domain.

%!TEX root = 0_main_TIG_validity.tex
\section{Empirical Setup} \label{empstudy}

\subsection{Research Questions}

In our study, we compare different TIGs in terms of their ability to generate valid artificial inputs, according to both automated validators and human assessors, and to preserve the original labels. As a secondary objective, we evaluate also the degree to which automated input validation matches human judgement.

\noindent\textbf{RQ1 (Comparing TIGs through automated validators):}  \emph{Which TIG generates more valid misbehaviour-inducing inputs, according to automated input validators?}

TIGs are more useful when the generated misbehaviour-inducing inputs are valid, whereas invalid inputs may contribute to a wrong assessment of the DL system quality. Automated input validation techniques are used to automatically filter out invalid inputs. In this research question, we compare the effectiveness of different TIGs in producing valid inputs, according to automated validators. %\vincenzo{do we need statistical significance tests for comparisons in RQ1, RQ2, RQ3?} \paolo{Yes: given two tools to compare T1, T2, let us build the 2x2 contingency matrix:  [[\#valid\_T1, \#invalid\_T1], [\#valid\_T2, \#invalid\_T2]]: Fisher test will gives us the $p$-value. We can take T1 = best TIG; T2 2nd best, 3rd best, etc. T1 can be boldface; the first with low $p$-value can be underscored. An alternative would be to use the Tukey's test (parametric) or the Dunn's test (non parametric), which return groups of techniques with high p-value separated from groups significantly worse, without requiring the comparison of each pair of techniques.}

\textit{Metric:} For each considered TIG, we measure the ratio of inputs that are considered valid by automated validators over all the generated inputs. 

\noindent\textbf{RQ2 (Comparing TIGs through human validators):}  \emph{Which TIG generates more valid misbehaviour-inducing inputs, according to human assessors?}

Automated validators offer a useful proxy for input validity. However, a more precise validity assessment of artificial inputs can only be provided through human judgement.

\textit{Metric:} For each considered TIG, we measure the ratio of inputs that are considered valid by human assessors over all the generated inputs. 

\noindent\textbf{RQ3 (Comparing TIGs in terms of label preservation):}  \emph{Which TIG generates more valid inputs that preserve the seed label?}

TIGs produce artificial inputs for which the predicted label differs from the expected one, i.e., usually the same class as the one of the input seed. 
In addition to assessing the validity of the artificial inputs, testers must verify whether these inputs actually preserve the expected label and, thus, the reported misbehaviours correspond to \textit{true misclassifications}.

\textit{Metric:} For each considered TIG, we measure the ratio of valid, label-preserving inputs, i.e., valid inputs that preserved the ground-truth label, over all the generated valid inputs. To assess validity and assign a ground-truth label to the generated inputs, we resort to human judgement.

\noindent\textbf{RQ4 (Automated vs human validity):}  \emph{Does automated validation match human validation?}

Automated input validation techniques are useful to filter out invalid inputs from artificially generated test suites with reduced human effort. However, automated validators are reliable only if human assessors agree with their outcome.

\textit{Metric:} To assess to what extent automated and human validators agree, we compare their respective validity assessments performed on the same artificial inputs, by checking if the inputs considered valid or invalid by an automated validator are confirmed to be so by a pair of human assessors, who act as our ground truth. 

Therefore, we measure the automated validators' \textit{Accuracy}, i.e., their ratio of correct validity predictions, as follows:

\begin{equation}
\scalebox{.9}{$\displaystyle Accuracy = \frac{TP + TN}{TP + TN + FP + FN}$}
\end{equation}

\noindent where: \textit{True Positives (TP):} valid for automated and human validators; \textit{True Negatives (TN):} invalid for automated and human validators; \textit{False Positives (FP):} valid for automated validators, invalid for humans; and \textit{False Negatives (FN):} invalid for automated validators, valid for humans.

\subsection{Subject Datasets and DL Systems}

We consider three popular image datasets: Modified National Institute of Standard and Technology (MNIST)~\cite{LecunBBH98}, Street View House Numbers (SVHN)~\cite{Netzer2011}, and ImageNet-1K~\cite{Deng2009}. These datasets are increasingly difficult classification tasks. In particular, ImageNet-1K is a quite challenging image classification task, with 1k classes and large-size images, representing a complex target for both TIGs and automated validators. For each of these tasks, we consider a corresponding pre-trained DL model that is (1) well-performing in terms of classification accuracy, (2) popular, and (3) widely adopted in the DL testing research. 

\noindent\textbf{MNIST} consists of $70\,000$ greyscale images of handwritten digits. The images are quite small ($28 \times 28$), while their pixel levels range from $0$ to $255$. Its classification task is simplified by the fact that all the digits are centred and have a black background. Due to its simplicity, this dataset is the most widely used to take the first steps as DL developers~\cite{keras-tutorial} %\footnote{\url{https://www.tensorflow.org/datasets/keras_example}} 
and rapidly assess prototypal solutions for testing DL systems~\cite{KimFY19, RiccioFSE20}. As DL classifier, we chose the same LeNet convolutional neural network~\cite{LecunBBH98} adopted by several DL testing works~\cite{PeiCYJ17, guo2018dlfuzz, DolaICSE21}. Its architecture is composed by two convolutional layers, followed by a fully connected layer. In particular, we used the weights of the model trained by Pei et al.~\cite{PeiCYJ17}.
%We used the trained classifier's weights by Pei et al.~\cite{PeiCYJ17}

\noindent\textbf{SVHN} contains $600\,000$ real-world images representing digits (from $0$ to $9$) cropped from house number plate pictures. These images are coloured (i.e., encoded as three RGB channels) and slightly bigger than MNIST ($32 \times 32$). Classifying SVHN images is more difficult than MNIST ones since the collected house number images occasionally include nearby digits and, unlike MNIST, the background and the digit do not have fixed colours. We used the \textit{All-CNN-A} architecture by Springenberg et al.~\cite{springenberg2014striving} which consists of $7$ convolutional layers and the weights provided by Dola et al.~\cite{DolaICSE21}.

\noindent\textbf{ImageNet-1K} is a large image database containing $1\,431\,167$ real-world images belonging to $1\,000$ non-overlapping classes. %, corresponding to as many WordNet~\cite{miller1998wordnet} nouns. 
This database is extremely popular since it has been used for the annual ImageNet Large Scale Visual Recognition Challenge (ILSVRC)~\cite{ILSVRC15} since 2012. ImageNet inputs have a higher resolution than the previously mentioned datasets (most classifiers use $225 \times 225$ resolution) and represent real-world classes, e.g., ``cat''  or ``pizza''. For this dataset, we used the pre-trained VGG16 deep convolutional neural network~\cite{simonyan2014very} ($13$ convolutional and $3$ fully connected layers) provided by the Keras library~\cite{keras-vgg}.%\footnote{\url{https://keras.io/api/applications/vgg}}.

Since some TIGs are developed with TensorFlow-Keras\footnote{\url{https://www.tensorflow.org}} (i.e., DeepXplore, DLFuzz, and DeepJanus), while the others use PyTorch\footnote{\url{https://pytorch.org}}, the same pre-trained DL classifiers cannot directly work with all the considered TIGs. We translated the original TensorFlow DL architectures to PyTorch  and copied the same weights in order to test functionally equivalent DL systems with all the TIGs and, thus, conduct a fair comparison.

%The MNIST system recognises handwritten digits from the MNIST dataset [42]; hence, it performs a classification task. Its DNN predicts which digit is represented in a greyscale image. In particular, we consider the popular convolutional DNN architecture provided by Keras [14].We trained this DNN on the MNIST training set using its default configuration, i.e., 12 epochs, batches of size 128, and a learning rate equal to 1.0. Our digit classifier achieved 99.8% classification accuracy on the MNIST testing set.

\subsection{Automated Input Validators Setup}

We considered the automated input validators presented in \autoref{validity}, DAIV and SelfOracle. To make these validators distribution-aware, we trained their VAE components on the same training set used to train the DL system under test. Then, we computed the respective reconstruction fidelity (DAIV) and reconstruction loss (SelfOracle) thresholds. 

As regards DAIV, we considered the same anomalous datasets as proposed in the original paper for MNIST and SVHN, i.e., FashionMNIST~\cite{xiao2017fashion} and Cifar-10~\cite{krizhevsky2009learning}, respectively. Since ImageNet-1K was not considered by the authors of DAIV, we applied their recommendations for choosing an anomalous dataset (see \autoref{dola}) and adopted the popular CelebA\footnote{We preprocessed CelebA images to have the same size as Imagenet-1K.} dataset~\cite{liu2015deep}. For SelfOracle, we adopted a low rate of false alarms ($0.01\%$) since most of the inputs in the original test sets represent nominal data by construction \footnote{Additional results obtained with SelfOracle for different thresholds are reported in the replication package.}.

%\vincenzo{Here I think we do not have enough space to mention that we had to develop the script to automatically compute the threshold for DAIV that was not present in the replication package.}
%\paolo{Yes, it's a detail that is not necessary in the paper, but people will find the script in the replication package.}

The VAEs of the two techniques share very similar architectures, with the exception of the different function being optimised (i.e., reconstruction fidelity maximisation for DAIV vs reconstruction loss minimisation for SelfOracle) and the different decoders' last layer, since DAIV outputs a probability distribution while SelfOracle returns the reconstructed image.

In particular, for MNIST and SVHN we reused, for both validators, the VAE architectures from the DAIV paper, i.e., a fully connected network for MNIST and a convolutional network for SVHN. Since ImageNet was not considered by the authors of DAIV, we adopted an architecture similar to the one used for SVHN but with a larger number of layers and nodes to effectively reconstruct more complex images. %\vincenzo{In this section, I have been a bit vague. I can use more space to report more details on the architecture, the final losses computed on the test set, and the thresholds. Also, the Imagenet DAIV architecture is slightly different from SelfOracle since it did not fit the memory (last layer has 2 times the weights of SelfOracle due to the different output.)} \paolo{If space allows, we can add these details, although they are not strictly necessary for the paper.}

\subsection{Compared Test Input Generators}

We considered the five TIGs described in~\autoref{tig} since they (1) represent diverse test generation approaches (i.e., RIM, GDLM, and MIM) and (2) are open-source and, thus, can be adapted to our needs. We applied the following modifications to ensure a fair and complete comparison.

\textbf{DeepXplore (DX) and DLFuzz (DLF)} originally covered only MNIST and ImageNet-1K. They have been extended thanks to the code of the DeepXplore variant developed by Dola et al.~\cite{DolaICSE21}, which also considers the SVHN dataset. Moreover, DeepXplore has been slightly modified to find a misclassification for a single DL system under test, rather than a differential behaviour among multiple DL systems as in the original code. 

\textbf{Sinvad (SV)} %is a TIG leveraging generative DL models and 
originally covered only MNIST and SVHN. We adapted this TIG to ImageNet-1K by integrating the pre-trained BigGAN~\cite{brock2018large} by Brock et al., which produces high-quality images and is adopted also by the Feature Perturbations generator~\cite{DunnISSTA21}. 

\textbf{Feature Perturbations (FPT)} originally covered only MNIST and ImageNet-1K. However, the GAN architecture adopted for MNIST was  compatible with other image datasets and, thus, we could effectively re-train it for SVHN. 
%\vincenzo{We can exclude the following sentence since, even if this functionality was not in the paper, there was something very similar in the code: ``Moreover, we modified the test generation task to consider all misclassifications, rather than misclassifications targeted to a specific class, as originally proposed by Dunn et al.''}

\textbf{DeepJanus (DJ)} was originally applied only to one image classification task, i.e., MNIST. We applied the same vectorial model representation proposed for MNIST also to SVHN, since they both model digits. However, we did not find an effective model-based representation for complex images and, thus, we did not apply DeepJanus to ImageNet-1K.

\subsection{Experimental Procedure}

To generate misbehaviour-inducing inputs, we ran the TIGs under comparison in the configurations suggested by the authors of the respective original papers.

We generated same-size test suites with each tool, i.e., $100$ inputs for MNIST/SVHN, $20$ inputs for ImageNet-1K (the smaller size of the latter is due to the complexity of the ImageNet task). We did not limit the TIGs' generation budget since we focused on input validity, rather than test generation efficiency and, thus, we needed a significative number of tests for each tool.
We fixed the same ground-truth label for each test generation problem, i.e., ``digit 5'' for MNIST/SVHN and ``pizza'' for ImageNet-1K, to limit the computational cost of our experiments, which remained anyway extremely high. 

The generated inputs were analysed both by automated validators (i.e., DAIV and SelfOracle) and human assessors. In this way, we collected information about the validity of each artificial input. 

We relied on a crowdsourcing platform to access a diverse and independent pool of human assessors~\cite{BehrendSMW11}. In the software engineering field, crowdsourcing is particularly useful for automating small tasks that can only be performed by humans~\cite{MaoCHJ17}, in exchange for an adequate remuneration~\cite{PastoreMF13}. In particular, we chose the Amazon Mechanical Turk platform\footnote{\url{https://www.mturk.com}} since it is easy-to-use, highly customisable and widely adopted to gather qualitative feedbacks~\cite{KitturCS08, HeerB10, RiccioFSE20}. 
To determine whether the artificially generated inputs belong to validity domain of the corresponding problem, we asked humans to classify the images generated by each tool. For each image, we asked which class was represented within the problem domain or if the image did not belong to that domain (i.e., it was invalid). For MNIST/SVHN, we listed all the $10$ possible classes along with the ``Not a handwritten digit/house number'' option. For ImageNet-1K (which has $1\,000$ classes), the user could choose between the expected class, the class actually predicted by the tested DL system, or the ``Another real-world object'' and ``No real-world objects'' options.

We adopted two solutions to ensure the quality of the human assessors' answers: (1) we added an attention check question (ACQ) to each survey and (2) we restricted the participation to workers with high reputation (above 95\% approval rate)~\cite{PeerVA14}. The ACQ consisted of an image for which the human choice is trivial. Therefore, we only accepted answers from users that passed the ACQ.

For each classification problem, we published surveys made of multiple questions, always including one ACQ. Each survey featured images from all TIGs, while each image was featured only in one survey. Each survey was answered by $2$ assessors, so that each input was assessed by $2$ human evaluators. In total, our human assessment consisted of $110$ surveys and involved $220$ human assessors.

Finally, we measured the number of answers in which both human assessors agreed in considering the analysed image as valid (i.e., assigned a class within the problem domain) or invalid, whereas we discarded the inputs on which the two assessors disagreed. 

%The Amazon Mechanical Turk platform allows us to measure the time spent to answer each question, which is necessary to answer RQ5.
%TODO: if we keep RQ5, I should mention that mturk allows us to measure the time spent to answer.

We determined the statistical significance of our conclusions by performing the Fisher's exact test~\cite{Fisher06}. %on the experimental results. 
In particular, we compared each pair of TIGs on the validity of their generated inputs (number of valid vs invalid inputs; RQ1 and RQ2) and on their label preservation capability (number of preserved vs non preserved labels; RQ3). 
A $p$-value $<\alpha$ = 0.05 (where $\alpha$ is the probability of a Type I error) was considered an indicator of statistical significance of the difference between the two compared TIGs. 
%Moreover, when $\lvert 1 - $ odds ratio $\rvert\gg0$, the expected relative proportion of valid vs invalid (true misclassifications vs false alarms for RQ3) is much lower for a TIG than the compared one.
As regards RQ4, we adopted the same statistical test to compare the results from our human study with each automated validator's assessment (considering the respective number of valid vs invalid inputs). In this case, $p$-value $\geq \alpha$ and a \textit{negligible} effect size (odds ratio lower than the conventionally accepted threshold, i.e., $1.43$) were considered as lack of statistically significant differences, i.e., automated validity was considered to match human judgement.

%!TEX root = 0_main_TIG_validity.tex
\section{Results} \label{results}

%!TEX root = 0_main_TIG_validity.tex
\begin{table}[]
\caption{RQ1-2-3: Comparison between TIGs in terms of generated valid misbehaviour-inducing inputs according to automated and human validators, and of preserved ground-truth labels. The best results are reported in bold. The underlined values are not statistically different from the best.} %\paolo{TODO: %(1) remove underline from highest values; 
%(2) check odds ratio and remove underline when not negligible}}
%; (3) report one decimal point for all values, not just for some; (4) report percentages for Human, not absolute values.}}

\label{tab:rq123}
\begin{center}
\begin{tabular}{llrrrrr}
\thickhline                                                                   
\multicolumn{1}{l}{\multirow{2}{*}{Dataset}} & \multicolumn{1}{c}{\multirow{2}{*}{Tool}} & \multicolumn{3}{c}{\% Valid} & \multirow{2}{*}{\begin{tabular}[c]{@{}r@{}}\% Hum. \\ Agree. 
\end{tabular}} & \multirow{2}{*}{\begin{tabular}[c]{@{}r@{}}\% Pres. \\ Labels 
\end{tabular}} \\
\multicolumn{1}{l}{}               	& \multicolumn{1}{c}{}     & DAIV     					& SO     				& Human    &                                                       \\ \hline
\multirow{6}{*}{MNIST}         	& DX       			      & 35       				        & 35     				& 81       			& \underline{93} & 92                                    \\
                                             	& DLF     			      & 37       				        & 85     				& \textbf{100}  		& \textbf{99} & \textbf{99}      		 \\
                                             	& SV       			      & \textbf{97}          			& \textbf{100}  			& \textbf{100}       	& \underline{96} & 58                                    \\
                                             	& FPT       			      & \underline{90}			       		        & \textbf{100} 			& \underline{99}       	& \underline{94} & 54                                    \\
                                             	& DJ       			      & \textbf{97}       			        & \textbf{100}  			& \textbf{100}       	& \underline{96} & 93                  \\[2pt] \thickhline
%\hline
%                                             	& Avg                              & 71         					& 84       				& 96         			& 96 & 79                                     \\ \thickhline

\multirow{6}{*}{SVHN}          	& DX                               & 51       					& \underline{99}     		& 77       			& \underline{68} & \textbf{79}       		  \\
                                             	& DLF                             & \textbf{100}      				& \textbf{100}  			& \textbf{96} 		& \underline{73} & 50                                     \\
                                             	& SV                               & \textbf{100}      				& \textbf{100}  			& \underline{90}       	& \underline{74} & 9                                       \\
                                             	& FPT                               & \underline{99}       			& \textbf{100}  			& 81			       	& \underline{75} & 46                                     \\
                                             	& DJ                               & 0        					& 1     				& 61       			& \textbf{76} & 9                                       \\[2pt] \thickhline
%\hline
%                                             	& Avg                             &   70       					& 80       				&  81        			&  73 & 38                                      \\ \thickhline

\multirow{5}{*}{ImageNet-1K}    & DX                              & \textbf{100}      				& \textbf{100}  			& \underline{90}       	& \textbf{100} & \underline{94}                    \\
                                             	& DLF                            & \textbf{100}      				& \textbf{100}  			& \textbf{100}  		& \textbf{100} & \underline{95}                    \\
                                             	& SV                              & \textbf{100}      				& \textbf{100}  			& 60       		        & 50 & \underline{83}                    \\
                                             	& FPT                              & \textbf{100}      				& \textbf{100}  			& \textbf{100}  		& \textbf{100} & \textbf{100}          		   \\[2pt] 
%\hline
%                                             	& Avg                             & 100      					& 100  				&   87       			& 88 & 93      		   		   \\
\thickhline                                                                   
\end{tabular}
\end{center}
\end{table}
%!TEX root = 0_main_TIG_validity.tex
\begin{table*}[]
\caption{RQ4: Automated validity vs ground truth (human validity). \textit{TP:} valid for automated and human validators; \textit{TN:} invalid for automated and human validators; \textit{FP:} valid for automated validators, invalid for humans; \textit{FN:} invalid for automated validators, valid for humans. Underline indicates no statistically significant disagreement in terms of accuracy.}

\label{tab:rq4}
\begin{center}
\begin{tabular}{ll
>{\columncolor[HTML]{C0C0C0}}r 
>{\columncolor[HTML]{C0C0C0}}r 
>{\columncolor[HTML]{C0C0C0}}r 
>{\columncolor[HTML]{C0C0C0}}r 
>{\columncolor[HTML]{C0C0C0}}r 
rrrrr
>{\columncolor[HTML]{C0C0C0}}r 
>{\columncolor[HTML]{C0C0C0}}r 
>{\columncolor[HTML]{C0C0C0}}r 
>{\columncolor[HTML]{C0C0C0}}r 
>{\columncolor[HTML]{C0C0C0}}r }
\thickhline
                            &                        & \multicolumn{5}{c}{\cellcolor[HTML]{C0C0C0}MNIST}                                                                                                                                                           & \multicolumn{5}{c}{SVHN}           & \multicolumn{5}{c}{\cellcolor[HTML]{C0C0C0}ImageNet-1K} \\
\multirow{-2}{*}{Validator} & \multirow{-2}{*}{TIG}  & TP                                             & FP                                             & FN                                             & TN                                            & Acc (\%) & TP   & FP  & FN   & TN  & Acc (\%) & TP      & FP       & FN      & TN     & Acc (\%)     \\ \hline
                & DX    & 10      & 18       & 65   & 0    & 11     
                            & 21   & 6   & 31   & 10  & 46     
                            & 18      & 2        & 0       & 0      & 90           \\
                & DLF  & 40   & 0   & 59   & 0  & 40     
                		   & 70   & 3   & 0    & 0   & 96     
                		  & 20      & 0        & 0       & 0      & \underline{100}          \\
                & SV   & 94      & 0   & 2    & 0    & \underline{98}     
                		  & 67   & 7   & 0    & 0   & 90     
                		  & 6       & 4        & 0       & 0      & 60           \\
               & FPT    & 85   & 1   & 8   & 0   & 91     
               		  & 60   & 14  & 1    & 0   & 80       
		  	  & 20      & 0        & 0       & 0      & \underline{100}          \\
              & DJ     & 92      & 0   & 4   & 0  & \underline{96}     
              		 & 0    & 0   & 46   & 30  & 39     
		 	 & -       & -        & -       & -      & -            \\ [2pt]
\hline
\multirow{-6}{*}{DAIV}      & Avg    & 64   & 4  & 28   & 0     & 67     
						   & 44 & 6   & 16 & 8   & 70     
						   & 16      & 1      & 0       & 0      & 87         \\ \thickhline

		& \multicolumn{1}{l}{DX} & \multicolumn{1}{r}{\cellcolor[HTML]{C0C0C0}15} & \multicolumn{1}{r}{\cellcolor[HTML]{C0C0C0}18} & \multicolumn{1}{r}{\cellcolor[HTML]{C0C0C0}60} & \multicolumn{1}{r}{\cellcolor[HTML]{C0C0C0}0} & 16     
& 51   & 16  & 1    & 0   & 75       
& 18      & 2        & 0       & 0      & 90           \\
                & DLF   & 99   & 0   & 0   & 0   & \underline{100}      
                		    & 70   & 3   & 0    & 0   & 96     
		            & 20      & 0        & 0       & 0      & \underline{100}          \\
                & SV    & 96    & 0   & 0  & 0   & \underline{100}      
                		   & 67   & 7   & 0    & 0   & 90     
		   	   & 6       & 4        & 0       & 0      & 60           \\
                & FPT    & 93   & 1   & 0   & 0    & 99     
                		   & 61   & 14  & 0    & 0   & 81     
		   	  & 20      & 0        & 0       & 0      & \underline{100}          \\
                & DJ   & 96    & 0   & 0    & 0   & \underline{100}      
                		 & 0    & 1   & 46   & 29  & 38     
		 	 & -       & -        & -       & -      & -            \\ [2pt]
\hline
\multirow{-6}{*}{SelfOracle}        & Avg  & 80   & 4     & 12  & 0    & 83       
							 & 50 & 8 & 9  & 6 & 76     
							 & 16      & 1      & 0       & 0      & 87        \\
\thickhline
\end{tabular}
\end{center}
\end{table*}

\begin{figure}[t!]
	\center \includegraphics[width=215pt]{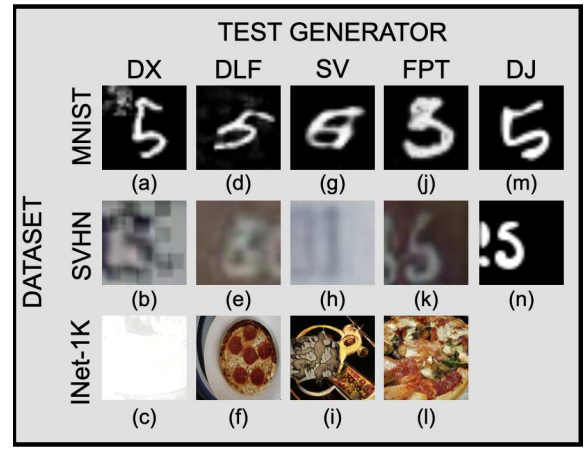}
        \caption{Selection of interesting inputs generated by TIGs, considered in the presentation of the results}
        	\label{fig:inputs}
\end{figure}

\subsection{RQ1: Comparing TIGs through Automated Validators}

The third and fourth columns of \autoref{tab:rq123} report the percentage of valid inputs, according to the automated validators. 

For MNIST, the techniques using generative DL models (i.e., SV and FPT) and the model-based DJ produced a significantly higher number of valid inputs than DX and DLF. %(p-value $<0.05$; low odds ratio). 
In fact, all images from the original MNIST dataset feature a black background and a greyscale digit in the centre. This property is either learned by generative ML models or preserved by the DJ model manipulation, which modifies the vector modelling the digit. Instead, DX and DLF can perturb all the image pixels, often generating images with non-completely black background (as shown in \autoref{fig:inputs}.a). Therefore, DX and DLF generated images with properties totally absent in the original dataset (on which the validators' VAEs were trained) and, thus, led to high reconstruction loss values (SelfOracle) and low reconstruction fidelity values (DAIV).

For SVHN, almost all tests ($\geq99\%$) generated by GDLM techniques and DLF are valid, according to both automated validators. %In particular, SV and DLF produce 100\% of valid inputs. 
For the other RIM technique (DX), there was disagreement between the validators: SelfOracle classified almost all the generated inputs as valid (i.e., 99\%), while for DAIV nearly half of them were valid (i.e. 51\%). This disagreement is mainly due to the different training procedure of their VAEs. In fact, almost all SVHN inputs generated by DX have characteristics that are present in 99.99\% of the training inputs%(and, thus, produce low reconstruction loss)
, but DAIV deemed half of them as more likely to belong to CIFAR-10, the anomalous dataset (e.g., for \autoref{fig:inputs}.b). DJ produced only $1$ valid input according to SelfOracle (none according to DAIV). DJ adopts a vectorial representation similar to the MNIST one and, thus, produces black and white images that preserve the digits' contours (see \autoref{fig:inputs}.n). Such images generated by DJ are very different from the SVHN dataset used for training the VAE of validators, which then classify these inputs as invalid.

For ImageNet-1K, all RIM and GDLM techniques generated only valid inputs, according to both automated validators. For this feature-rich dataset, the VAE of automated validators learned to reconstruct all the generated artificial inputs, even when they are challenging to classify (see \autoref{fig:inputs}.i), because they learn a huge amount of low-level image features from the extremely rich ImageNet-1K dataset and they can reconstruct almost any image using such low-level features.%for the DL systems under test.

\begin{tcolorbox}
\textbf{Summary RQ1}: \textit{All RIM and GDLM techniques can produce valid inputs for all datasets, according to automated validators. Sinvad always generated the highest percentage ($\geq97\%$) of valid inputs.}
\end{tcolorbox}

\subsection{RQ2: Comparing TIGs through Human Validators} \label{sec:rq2}

The fifth column of \autoref{tab:rq123} shows the percentage of valid inputs %generated by each tool on each subject
 according to human validators, after removing the inputs on which the assessors disagreed, while the sixth column reports the percentage of inputs on which there was agreement among human validators. %\paolo{We should report also the percentage of inputs on which there was disagreement: I'd add a column "\% Dis", "Human"}

For MNIST, human assessors judged the vast majority of artificially generated images as belonging to the validity domain ($96\%$ on average). 
DLF, SV, and FPT generated the highest percentage of valid inputs, which was significantly higher than DX %(p-value $<0.05$; low odds ratio) 
and comparable to FPT. %On the other hand, DX achieved the worst performance (i.e., 18\% invalid inputs, while for 7\% assessors did not agree), probably because of the aggressiveness of the \textit{light} operator that transformed some seeds in almost completely white images, that were not recognisable anymore by humans.

For SVHN, the average percentage of valid inputs was $15\%$ lower than for MNIST. This is probably because SVHN images represent complex images at low resolution and thus can be challenging for human assessors to recognise (e.g., \autoref{fig:inputs}.b). DLF generated a higher percentage of valid inputs than the other techniques (i.e., 96\%), performing significantly better than DX, FPT, and DJ. %(p-value $<0.05$; low odds ratio).

For ImageNet-1K, both FPT and DLF generated only valid inputs. Instead, only 60\% of SV inputs were deemed valid. Moreover, human assessors reached an agreement only for half of the SV inputs. 

%(the remaining 70\% is actually split between 20\% clearly invalid inputs and 50\% on which the assessors did not agree). %(e.g., \autoref{fig:inputs}.i is an invalid input) %This result is probably due to the fact that SV (unlike FP) does not constrain the generative models' activation values within the ones observed during training, thus generating inputs that are invalid or challenging to recognise in complex input spaces.

\begin{tcolorbox}
\textbf{Summary RQ2}: \textit{According to human assessors, all techniques can produce valid inputs for all datasets. By introducing slight pixel perturbations, DLFuzz generated the highest percentage ($\geq96\%$) of valid inputs for all subjects.}
\end{tcolorbox}

\subsection{RQ3: Comparing TIGs in Terms of Label Preservation}

The last column of \autoref{tab:rq123} reports the ratio of preserved labels, i.e., the percentage of valid (and by construction misclassified) inputs that preserved their ground-truth label, hence resulting in a true misclassification, among those on which there was no disagreement between the $2$ human assessors.

For MNIST, DLF achieved the significantly best label preservation ratio (i.e., 99\%).
%, performing comparably to DJ %(p-value $\geq0.05$) 
%and significantly better than the remaining TIGs %(p-value $<0.05$; low odds ratio). 
On the other hand, GDLM techniques (i.e., FPT and SV) showed a label preservation ratio lower than the other TIGs, despite generating a high number of valid inputs. \autoref{fig:inputs}.g,j shows GDLM inputs that changed their ground-truth label.

For SVHN, DX achieved a label preservation ratio significantly higher than the other TIGs. %, although only half of its inputs were classified as valid by human assessors. 
Instead, SV and DJ showed a dramatically low percentage of label preservations (i.e. $9\%$). As regards SV, this result is particularly interesting since it achieved the second highest percentage of valid inputs according to human assessors (see~\autoref{sec:rq2}). Therefore, we can conclude that SV was able to generate images looking like real house numbers, but it also transformed their original ground-truth label into another one (e.g., \autoref{fig:inputs}.h is a ``1'' for human assessors).

For ImageNet-1K, all TIGs achieved comparable label preservation ratios, %(p-value $\geq0.05$), 
always $>83\%$. In particular, all inputs generated by FPT were both deemed  valid by humans and preserved their ground-truth label (e.g., \autoref{fig:inputs}.l).

\begin{tcolorbox}
\textbf{Summary RQ3}: \textit{The valid inputs generated by TIGs did not always preserve their ground-truth label. This result was particularly unfavourable when generating SVHN tests, where on average only $38\%$ valid inputs were label-preserving.}
\end{tcolorbox}

\subsection{RQ4: Automated vs Human Validity}

\autoref{tab:rq4} reports the metrics to assess the agreement between automated and human validators. In particular, the top rows consider the DAIV validator, while the bottom of the table shows the results for SelfOracle.

For MNIST, both automated validators achieved $\geq90\%$ accuracy for SV, FPT, and DJ. 
%$3$ out of $5$ TIGs.  
Instead, SelfOracle showed a dramatically higher accuracy than DAIV on DLF inputs, by reducing the number of false negatives from $59$ to $0$ (e.g., \autoref{fig:inputs}.d is a false negative for DAIV and a true positive for SelfOracle). Overall, SelfOracle achieved a higher accuracy than DAIV ($+16\%$). Both automated validators strongly disagreed with human assessors on DX inputs' validity. In fact, nearly $2$ out of $3$ DX inputs were classified as invalid by the automated validators, while they were judged as valid by human assessors (see column FN).

For SVHN, both automated validators achieved $\geq70\%$ accuracy across all TIGs. Also on this dataset, SelfOracle showed a higher accuracy than DAIV ($+6\%$). The automated validators achieved similar accuracies on all TIGs but DX, on which accuracy is $29\%$ higher for SelfOracle than DAIV. In fact, DAIV has $30$ more false negatives (i.e., it classified inputs  that were valid for humans, like \autoref{fig:inputs}.b, as invalid) than SelfOracle. Both automated validators disagreed with human assessors on SVHN images generated by DJ. The high number of false negatives suggests that, despite the inputs generated by DJ differ from the ones in the dataset, humans can still recognise house numbers within these images (see \autoref{fig:inputs}.n).

For ImageNet-1K, the automated assessors totally agreed among them for all inputs, achieving a statistically significant match with humans on $2$ out of $4$ TIGs with $100\%$ accuracy i.e., DLF and FPT. As regards SV, the automated validators showed a lower accuracy ($60\%$) since the generated images were in-distribution for the VAEs, but were judged by humans as not real-world images (e.g., \autoref{fig:inputs}.i). Indeed, the low-level features used to create these images belong to the training set, so the result was deemed in-distribution by the automated validators, but the high-level result has no meaning to humans, being regarded as an incoherent patch of disparate elements. 

\begin{tcolorbox}
\textbf{Summary RQ4}: \textit{Automated validators' assessment showed a good match with human judgement ($78\%$ accuracy). SelfOracle achieved higher accuracy than DAIV across all considered datasets ($7\%$ higher).}%, on average)}
\end{tcolorbox}

\subsection{Threats to Validity}

\noindent\textbf{Internal Validity}: A comparison between TIGs for DL may introduce threats to internal validity if the system under test is trained separately in each TIG's specific framework (i.e., TensorFlow-Keras and PyTorch, in our case). To mitigate this threat, we used pre-trained TensorFlow models, translated their architecture to PyTorch, and used the same weights in order to preserve their functionality. Another threat could be introduced by survey participants, which might have provided random answers. We mitigated this threat by adding an ACQ to each survey and restricting the participation to workers with high reputation. 

\noindent\textbf{External Validity}: The choice of subject DL systems and datasets is a possible threat to the external validity. To mitigate this threat, we chose $3$ increasingly difficult classification tasks on popular datasets, including ImageNet-1K, which contains high-resolution images belonging to 1\,000 different classes. Moreover, we considered pre-trained  state of the art DL architectures that are widely used in the literature. In this work, we focused on image classification problems. To generalise our findings to regression problems, such as steering angle or eye gaze prediction, we could define misbehaviours based on a prediction error threshold~\cite{zhang2018deeproad, riccio2021deepmetis}. Given such a threshold, our study can be replicated in these domains.

\noindent To ensure the \textbf{Reproducibility} of our results, we make our experimental data available online. Moreover, we considered only open source tools and subjects in our evaluation~\cite{replicationPkg}.

\section{Lessons Learnt and Actionable Insights} \label{discussion}

\subsection{Imperceptible Pixel Perturbations Ensure Validity} DLF mostly generated valid inputs, according to humans, whereas, DX applied more aggressive pixel perturbations, achieving a worse performance. In particular,  DX's ``light'' operator transformed some seeds in almost completely white (or black) images, no longer recognisable by humans (e.g., \autoref{fig:inputs}.c).

To mitigate issues introduced by too aggressive pixel perturbations, RIM techniques could adopt a threshold over which the input would not be perturbed further. In particular, this threshold is defined on the distance (e.g., Euclidean distance) between the original input seed and the perturbed input. Instead, the considered RIM techniques do not consider any threshold or only define a distance threshold on single perturbations, that could still accumulate in a substantial (and invalidating) input modification.

\subsection{Pixel Perturbations Do Not Generate Novel Images, Unlike GDLM Techniques} RIM techniques introduce limited novel information, since they just modify seeds from the original datasets, differently from GDLM techniques, which introduce more novelty by sampling the latent space in a generative way.

TIG developers could use GDLM techniques to produce novel inputs, but at the same time to incorporate automated input validation to obtain inputs that are more likely to be valid. Then, they could submit such inputs as seeds to RIM techniques to generate valid misbehaviour-inducing inputs through imperceptible pixel perturbations.

\subsection{Generative DL Models Produce Unseen Inputs, but Should Carefully Explore the Latent Space} GDLM generate novel images, by sampling from the input distribution. Unsurprisingly, GDLM inputs are deemed as valid by automated validators, since these generators are distribution-aware. Albeit GDLM inputs possess similar features to the original dataset, they may still be invalid for humans or transform the original label into another one. 

Therefore, TIGs should carefully perturb latent vectors. In fact, SV achieved poor performance on SVHN in terms of label preservation, and on ImageNet-1K in terms of human validity. This result is probably due to the fact that SV (unlike FPT) does not constrain the generative models' activation values within the ones observed during training, thus generating inputs that are invalid or challenging to recognise for humans in complex input spaces.

\subsection{Model-Based Techniques Need High-Quality Model Input Representations} DJ mostly produced valid inputs and preserved ground-truth labels, when provided with a high-quality model representation, as for MNIST. Otherwise, it generated the lowest number of valid inputs (i.e., for SVHN). Moreover, input domain models are not always available, especially for complex domains (e.g., real-world pictures in ImageNet-1K).

\subsection{Distribution-Aware Automated Validators Struggle with Feature-Rich Datasets} On ImageNet-1k, both  automated validators deem all inputs as valid, failing to reject any input that is instead invalid for humans.
This result is due to the richness of the ImageNet-1K dataset, that contains a variety of high resolution real-world images from which the VAE-based validators learns several low-level features.
In fact,  SelfOracle VAE learns to reconstruct all the diverse features from the training dataset, which are found also in the artificially generated inputs. As regards DAIV, we configured it with an anomalous dataset that encodes different categories than ImageNet-1K (as suggested by its authors~\cite{DolaICSE21}), i.e., celebrity faces from CelebA. However, DAIV VAE can still reconstruct many anomalous images, since they share similar features with ImageNet-1K inputs, being both real-world images (although CelebA contains a lower variety of features than ImageNet-1K). As a result, the distributions of the reconstruction fidelity values for  nominal and anomalous datasets overlap substantially, making it difficult for DAIV to find an effective discriminative threshold. Indeed, anomalous datasets should contain  different features from the nominal dataset, in addition to encoding different categories, which might be difficult to achieve with feature-rich datasets. In conclusion, both DAIV and SelfOracle suffer from similar limitations, being dependent on the capability of the VAE, which in turn depends on the low-level image features present in the original training set.  However, DAIV also depends on the choice of the anomalous dataset, which might be difficult to make, especially for nominal feature-rich datasets, such as ImageNet-1K.

Therefore, we suggest to use DAIV when there is the availability of anomalous datasets that contain different features from the nominal dataset, in addition to having the same dimension and encoding different categories (as suggested by DAIV's authors).

\subsection{Humans Actually Check If Inputs Semantically Belong to the Domain, Whereas Automated Validators Check If Inputs Are in the Same Distribution of the Training Set}

Distribution-aware automated validators heavily depend on the low-level image features present in the original training set. As an example, the RIM inputs for MNIST may contain slight noise in the background that preserves the input semantic according to humans, but affects the automated validators decision, since the inputs from the training dataset have a completely black background. 

We believe that the creation of effective automated validators is still an open research problem that deserves future investigation, especially for feature-rich datasets. Our study suggests a few improvements for such future research. For instance: (1) train automated validators also on inputs augmented with imperceptible noise, to make the validators more robust; (2) incorporate semantic information, possibly collected from humans, to bridge the gap between in-distribution analysis and human validity.

\subsection{Label Preservation Is Often Overlooked by Current Testing Approaches} Mechanisms to ensure label preservation should be integrated into TIGs, along with validity monitors, to increase  usefulness of the generated test inputs. In fact, automated validators only check whether a test input is within the same distribution as the original dataset, but neglect the assessment of label preservation. %This introduces an overhead for human testers for re-classifying the artificially generated inputs.

%!TEX root = 0_main_TIG_validity.tex
\section{Related Work} \label{related}

\subsection{Test Adequacy for Testing DL Systems}
%\subsection{Input Validity in Software Testing}
%\subsection{Empirical Comparison of Test Input Generators for DL}

Several SE studies evaluated TIGs for DL, comparing them on various test effectiveness and adequacy metrics. %, mostly as part of their proposed technique's assessment.
The most basic metric is the number of exposed misbehaviours: a TIG is more effective than others if it can expose a higher number of misbehaviours~\cite{AbdessalemNBS18, zhang2018deeproad, GambiMF19}. Other studies  avoid rewarding TIGs that repeatedly expose too similar (or even the same) misbehaviours by considering also the misbehaviours' diversity, e.g., based on Euclidean distance in the input space~\cite{RiccioFSE20} or  coverage of the feature space~\cite{ZohdinasabISSTA2021, nguyen2021salvo, zohdinasabTOSEM2022}.

%While code coverage metrics are extensively adopted to measure test adequacy for traditional software, these metrics are not suitable to DL software since most of the DL systems’ behaviour does not depend on their code, but on other aspects such as their training data, their architecture or the trained weights. Therefore, 

Researchers have proposed specific test  criteria for DL systems. TIGs are often evaluated in terms of such DL specific adequacy metrics.

Pei et al.~\cite{PeiCYJ17} proposed the neuron coverage criterion, which counts the number of neurons activated by a test input. In particular, a neuron is considered activated if its output value is higher than a predefined threshold. DeepGauge~\cite{Ma-ASE-2018} extends neuron coverage with additional, finer-grained criteria that partition the neurons' activation values into bins to be covered. DeepCT~\cite{LeiJXLLLZ19} introduces, instead, combinatorial criteria that consider the interactions between neurons within a single layer or in adjacent layers. These neuron-based adequacy criteria are focused on covering specific sets of neurons or model layers. Despite neuron-based coverage criteria have been extensively used to evaluate TIGs~\cite{tian2018deeptest, guo2018dlfuzz, demir2019deepsmartfuzzer, xie2019deephunter, DolaICSE21}, empirical results showed that higher neuron coverage does not necessarily correlate with a higher number of exposed misbehaviours, while it may lead to the generation of less natural inputs~\cite{CanadaFSE20}.

Kim et al.~\cite{KimFY19, KimFY20} designed DL-specific test adequacy criteria based on the degree of ``surprise'' of an input for a given  neural network with respect to the training data. Then, they define a coverage criterion from the surprise measure by using bucketing, i.e., by counting how many of the $k$ surprise bins are covered by the test set.

Zhang et al.~\cite{ZhangICSE20} compared TIGs using alternative uncertainty metrics such as prediction confidence and variation ratio. They found  that the generated inputs follow common uncertainty patterns, but largely miss other combinations of uncertainty metric values, and that those uncommon patterns are harder to automatically defend from, when protecting a neural network from adversarial attacks.

% mutation
The statistical notion of mutation adequacy introduced by Jahangirova and Tonella~\cite{JahangirovaICST20} can be used to assess the TIGs' ability to expose DL model mutations, i.e., artificially injected faults that simulate real faults. TIGs guided by mutation adequacy~\cite{riccio2021deepmetis} can expose more real DL faults from the fault taxonomy by Humbatova et al.~\cite{HumbatovaICSE20} than TIGs guided by neuron coverage and prediction confidence.

Unlike our work, the empirical comparisons performed by the studies mentioned above do not take into account the notion of test input validity. They also overlook label preservation, which is taken for granted.

%high level and low level features?

\subsection{Validity Assessment of Test Input Generators}

Most of the techniques in the literature strive to generate tests within the validity domain by limiting themselves to slight, imperceptible perturbations to the inputs from the original training set. However, only a few works in the literature actually performed a validity assessment of their results.

Dola et al.~\cite{DolaICSE21} adopted an automated validity assessment based on VAEs. They showed that adversarial TIGs tend to produce a high number of invalid tests, which contribute to the increase of test adequacy metrics. Unlike our work, they limited the comparison to adversarial generators, which manipulate raw input data, and considered only validity assessment techniques dependent on an anomaly set. Instead, we considered also non-adversarial TIGs %that leverage either generative ML models or a model-based input representation
and the SelfOracle~\cite{StoccoICSE19, StoccoGAUSS20, StoccoJSEP21} validity assessment technique, which relies only on nominal data. Moreover, we compared the outcome of automated validators to human assessors' judgement and investigated the problem of ground-truth label preservation.

Few papers evaluated artificially generated inputs for DL by involving human assessors. 

Tian et al.~\cite{TianEMSE2021} showed, through human assessment, that traditional DL training strategies produce models that may make unreliable inferences, e.g. object classifications based on the surrounding environment, rather than the predicted object. 

Attaoui et al.~\cite{attaoui2022black} evaluated their feature extraction and clustering technique for DL systems, showing that smaller sets of artificial images (i.e., $5$ inputs) with common features are not worse than larger sets of images (i.e., $10-15$ inputs) for human assessors whose task is to detect the common root causes of DNN misbehaviours. 

Riccio and Tonella~\cite{RiccioFSE20} performed a human assessment of validity and showed that misbehaviour-inducing artificial inputs are more likely to be invalid when generated for a high quality DL system than for a low quality one. None of these works compared the generated inputs' validity according to humans nor compared human judgement with automated validity assessment techniques. 

Our study is the first to compare different kinds of TIGs, while evaluating the generated inputs both from the perspective of automated evaluators and human assessors. It is also the first to compare human vs automated validity assessment and the label-preservation issue.

%completely relied on automated validity assessment but did not check the validity according to human evaluators.

%Stocco et al.~\cite{} also adopted a VAE-based assessment to automatically detect out-of-distribution inputs.

%Few studies evaluated the artificially generated test inputs by conducting experiments involving human assessors. Riccio and Tonella~\cite{RiccioFSE20} showed that most of the failure-inducing test inputs for a high-quality DL model are invalid for humans.
%!TEX root = 0_main_TIG_validity.tex
\section{Conclusions and Future Work} \label{conclusion}

Our empirical study showed that TIGs for DL can generate, to different extents, valid inputs. %FPT, a tool leveraging generative DL models, stood out by producing $97\%$ of valid inputs, according to automated and human validators.
The root cause for input invalidity varies across different TIGs: RIM’s invalidity is caused by the corruption of a large number of pixels; GDLM’s invalidity by the lack of continuity in the latent space; MIM’s by the lack of a high quality input model representation.
%However
Furthermore, TIGs occasionally failed to preserve the expected label of the inputs (down to $38\%$ label-preserving inputs for the SVHN dataset). Automated validators matched well with human judgement ($78\%$ accuracy), but showed limitations with either simple datasets, such as MNIST, or feature-rich ones, such as ImageNet-1K.

Valid, label preserving artificially generated inputs obtained with our experimental procedure can be considered as a first step towards reliably evaluating multiple aspects of TIGs, such as input diversity/uniqueness or the features that characterise them. The lessons learnt in this work can be leveraged to improve existing TIGs and automated validators. In our future work, we plan to further generalise our results to a wider sample of DL systems, including industrial ones. 

%\textbf{Valid, label preserving artificially generated inputs obtained with our experimental procedure can be used to reliably evaluate multiple aspects of TIGs, such as their mutation killing ability.} \vincenzo{I would remove this since I do not want to spoil our follow-up extension}

%\section*{Acknowledgment}
%This work was partially supported by the H2020 project PRECRIME, funded under the ERC Advanced Grant 2017 Program (ERC Grant Agreement n. 787703). The authors would like to thank all the members of the PRECRIME team for their support and the useful discussions.

\bibliographystyle{IEEEtran}
\balance
\bibliography{biblio}

\end{document}